\documentclass[printer]{aa}
\usepackage{graphicx,txfonts,amssymb,natbib,epsfig}
\authorrunning{C. Fedeli, et al.}
\titlerunning{Strong cluster lensing in the {\sc MareNostrum Universe}}

\begin{document}

\title{Strong lensing in the {\sc MareNostrum Universe} II: scaling relations and optical depths}

 \author{C. Fedeli\inst{1,2,3}\thanks{E-mail: cosimo.fedeli@unibo.it}, M. Meneghetti\inst{2,3}, S. Gottl\"ober\inst{4}, G. Yepes\inst{5}}

\institute {$^1$ Dipartimento di Astronomia, Universit\`a di Bologna,
  Via Ranzani 1, I-40127 Bologna, Italy\\
  $^2$ INAF-Osservatorio Astronomico di Bologna, Via Ranzani 1, I-40127 Bologna, Italy\\
  $^3$ INFN, Sezione di Bologna, Viale Berti Pichat 6/2, I-40127 Bologna, Italy\\
  $^4$ Astrophysikalisches Institut Potsdam, An der Sternwarte 16, D-14482 Potsdam, Germany \\
  $^5$ Grupo de Astrof\'isica, Universidad Aut\'onoma de Madrid, Madrid E-28049, Spain}

\date{\emph{Astronomy \& Astrophysics, submitted}}

\abstract{The strong lensing events that are observed in compact clusters of galaxies can, both statistically and individually, return important clues about the structural properties of the most massive structures in the Universe. Substantial work is ongoing in order to understand the degree of similarity between the lensing cluster population and the population of clusters as a whole, with members of the former being likely more massive, compact, and substructured than members of the latter. In this work we exploit synthetic clusters extracted from the {\sc MareNostrum Universe} cosmological simulation in order to estimate the correlation between the strong lensing efficiency and other bulk properties of lensing clusters, such as the virial mass and the bolometric X-ray luminosity. We found that a positive correlation exist between all these quantities, with the substantial scatter being smaller for the luminosity-cross section relation. We additionally used the relation between the lensing efficiency and the virial mass in order to construct a synthetic optical depth that agrees well with the true one, while being extremely faster to be evaluated. We finally estimated what fraction of the total giant arc abundance is recovered when galaxy clusters are selected according to their dynamical activity or their X-ray luminosity. Our results show that there is a high probability for high-redshift strong lensing clusters to be substantially far away from dynamical equilibrium, and that $30-40\%$ of the total amount of giant arcs are lost if looking only at very X-ray luminous objects.}

\maketitle

\section{Introduction}\label{sct:introduction}

Explaining the process of cosmic structure formation is one of the major successes of the standard model for cosmology. Accordingly, tiny dark matter density fluctuations produced during the inflationary era grew up due to gravitational instability, eventually detaching from the overall expansion of the Universe and collapsing into bound structures. The observed statistical properties of these objects can help to understand the large scale matter distribution, and the way in which this has evolved during the history of the Universe. 

Gravitational lensing is one of the most powerful tools for studying the formation of cosmic structures in general, and the mass assembly history of the most massive ones, galaxy clusters, in particular. Weak gravitational lensing can be used to perform non-parametric reconstruction of the outer dark matter halo profile \citep{KA95.1,BA01.1,OG09.1}, while strong lensing allow to put constraints on the inner slope \citep{RI10.1}, that is one of the main theoretical predictions of the cold dark matter paradigm. The statistics of strong lensing images also is a particularly interesting tool for the study of structure formation, since it is sensitive to both the abundance of galaxy clusters and their average inner structure. It is hence expected to give important hints on the assembly history of cosmic structures as well as on the background cosmology \citep{BA03.1}.

Much work has been devoted in the past decade to the gravitational arc statistics, that is the abundance of strongly elongated images that are produced by galaxy clusters when acting as gravitational lenses on background galaxies. Early results pointed toward a deficit of observed arcs with respect to the theoretical predictions for a standard $\Lambda$CDM cosmology \citep{LE94.1,BA98.2,LU99.1}, which neither baryonic physics \citep{PU05.1,HI08.2,WA08.1}, nor substructure \citep{ME00.1,ME03.2} and cluster mergers \citep{FE06.1,FE07.1} seemed able to cure. Whether this discrepancy survives nowadays is unclear \citep{FE08.1}, however this issue propelled substantial effort in order to better understand the cluster population responsible for strong lensing events, and to what degree this ensemble differs from the overall cluster population \citep{HE07.1,ME10.1}. Moreover, considerable work has been done on the observational side as well, in order to explore the high-redshift tail of the cluster lens population \citep{GL03.1,ZA03.1}, and to pave the way toward machine-based gravitational arc finding \citep{LE04.1,SE07.1}.

The dependence of the cluster lensing efficiency on the inner structure, dynamical state and merger activity of individual objects has been thoroughly studied \citep{TO04.1,ME07.1}. An important issue that has been however scarcely investigated about strong lensing cluster is how the lensing efficiency correlates with bulk cluster properties, such as mass or X-ray luminosity. This aspect is extremely interesting, since it allows to better quantify the statistical properties of the cluster lens population, and in principle would allow to directly map the abundance of galaxy clusters into the arc abundance on the sky without intermediate calculations. In this work we tackle this issue making use of the large {\sc MareNostrum Universe} cosmological simulation. This simulation contains adiabatic gas in addition to dark matter, and hence allows to characterize the cluster lens population taking into account the role of baryonic physics on structure formation. We additionally estimated the optical depth of the whole cluster population, which allows to compute the total number of arcs expected in the sky, and how this depends on selection biases such as the dynamical state or the X-ray luminosity.

The rest of this paper is organized as follows. In Section \ref{sct:simulation} we summarize the properties of the {\sc MareNostrum Universe} simulation, and the calculations relevant for strong lensing of simulated clusters. Further details on this can be found in \citet{ME10.1}. In section \ref{sct:population} we summarize the redshift distributions of the strong lensing cluster population. In Section \ref{sct:results} we report our results and in Section \ref{sct:summary} we summarize our conclusions. In the remainder of this work, we refer to \citet{ME10.1} as to Paper I.

\section{Lensing in the {\sc MareNostrum Universe} simulation}\label{sct:simulation}

The {\sc MareNostrum Universe} \citep{GO07.1} is a large scale non-radiative SPH cosmological simulation performed with the publicly available code {\sc Gadget2} \citep{SP05.1}. The simulation box has a comoving side of $500$ Mpc $h^{-1}$ and contains $2\times 1024^3$ particles, equally subdivided in number between dark matter and baryons. The mass of each dark matter particle equals $8.24 \times 10^{9} M_\odot h^{-1}$, and that of each gas particle, for which only adiabatic physics is implemented, is $1.45 \times 10^{9} M_\odot h^{-1}$. The cosmological parameters of the simulation are in agreement with the WMAP-1 year data release \citep{SP03.1}, namely $\Omega_{\mathrm{m},0} = 0.3$, $\Omega_{\Lambda,0} = 0.7$ and $\sigma_8 = 0.9$ with a scale invariant primordial power spectrum. The baryon density parameter is set to $\Omega_{\mathrm{b},0} = 0.045$. How this parameter set deal with the later WMAP releases \citep{SP07.1,KO09.1,KO10.1} is discussed in Section \ref{sct:summary}. In the following we describe only the main issues of our strong lensing analysis, deferring further detail to Paper I.

Bound structures within the simulation at each redshift snapshot were identified with a FOF algorithm \citep{KL99.1} with a basic linking length of $0.17$ times the mean interparticle distance. All the FOF groups with mass larger than $10^{13}M_\odot h^{-1}$ were then stored into sub-boxes of cubic shape with side length $5$ Mpc $h^{-1}$ for the subsequent lensing analysis. For each structure, the three-dimensional density field was computed on a regular grid inside the relative sub-box, and then projected along the three orthogonal coordinate axes. Bundles of light rays were than traced through the central part of each plane \citep{ME07.1}, evaluating for each of them the deflection angle as the sum of the contributions given by each surface density cell. The final deflection angle map has a side length of $1.5$ Mpc $h^{-1}$, which allows to fairly capture the details of the critical line structure (if present).

A preliminary analysis, performed with low resolution deflection angle maps has shown that $49,366$ clusters were able of producing critical lines for sources at $z_\mathrm{S} = 2$ in at least one projection. For each of the projections of these clusters we computed high resolution deflection angle maps for the subsequent strong lensing analysis. We evaluated the cross sections $\sigma_{R_0}$ for arcs with length-to-width ratio $R\ge R_0$ by adopting the semi-analytic algorithm of \citet{FE06.1}. Such algorithm reproduces well the results of fully numerical lensing simulations, while being substantially faster. Moreover, being based only on the properties of the deflection angle map, it is equally applicable to analytic \citep{FE07.1} and to numerical lenses. For such computations the equivalent angular size of sources was kept fixed to $0.5$ arcsec, and the ellipticity thereof was randomly drawn in the interval $[0,0.5]$. Given the extremely large number of projections that we considered, we were forced to examine only one single length-to-width threshold, $R_0 = 7.5$, and one individual source redshift, that we set to $z_\mathrm{S} = 2$. These choices are rather common in strong lensing statistics studies,  with $R_0=7.5$ being a fair compromise between having good number statistics and strongly distorted images, and $z_\mathrm{S}=2$ being typical for many real strong lensing clusters (see, e.g., \citealt{EL07.1,LI08.1}). Of the $148,098$ ($=3 \times 49,366$) projections that we have analyzed, only $11,347$ produced a non-vanishing cross section. This number is however by far the largest ever considered for theoretical strong lensing studies to date.

The total number of arcs in the sky with $R\ge R_0$ that are produced by galaxy clusters that lens sources at a given redshift $z_\mathrm{S}$, and that are visible below some limiting magnitude cut $m_\mathrm{lim}$, is given by $\mathcal{N}_{R_0}(m_\mathrm{lim}) = 4\pi\, n(m_\mathrm{lim},z_\mathrm{S})\tau_{R_0}(z_\mathrm{S})$, where $n(m_\mathrm{lim},z_\mathrm{S})$ is the number of sources that, in the unit solid angle, have redshift $z_\mathrm{S}$ and magnitude below the limit $m_\mathrm{lim}$. Note that, in order for this estimate to be self-consistent, the number density has to be always corrected for the lensing magnification bias \citep{BA01.1,FE08.1}. This bias has the effect of increasing the angular density of visible sources if the intrinsic number counts are steep enough, while decreasing it if the number counts are relatively shallow.

The optical depth for long and thin arcs produced by sources at redshift $z_\mathrm{S}$ is instead defined as

\begin{equation}\label{eqn:tau}
\tau_{R_0}(z_\mathrm{S}) \equiv  \int_0^{z_\mathrm{S}} \int_0^{+\infty} N(M,z_\mathrm{L}) \frac{\bar{\sigma}_{R_0}(M,z_\mathrm{L})}{4\pi D^2_\mathrm{A}(z_\mathrm{S})} dM dz_\mathrm{L},
\end{equation}
where $N(M,z)$ represents the total number of structures present in the unit redshift around $z$ and in the unit mass around $M$. The function $D_\mathrm{A}(z)$ is the angular diameter distance out to redshift $z$, while $\bar{\sigma}_{R_0}$ is the average cross section of each cluster in the {\sc MareNostrum Universe}. This average is taken over the three orthogonal projections (see above). Please note that this is equivalent to the correct procedure of computing three different optical depths, using a different cluster projection for each one, and then average the result.

When only a finite discrete set of masses is available, as is the case here and in all the numerical or semi-analytic studies, then the Eq. (\ref{eqn:tau}) reduces to

\begin{equation}\label{eqn:numtau}
\tau_{R_0}(z_\mathrm{S}) = \int_0^{z_\mathrm{S}} \left[\sum_{i=1}^{\ell-1} \frac{\bar{\sigma}_{R_0,i}^*}{4\pi D^2_\mathrm{A}(z_\mathrm{S})} \int_{M_i}^{M_{i+1}} N(M,z_\mathrm{L}) dM \right] dz_\mathrm{L},
\end{equation}
where $\ell$ is the number of strong lensing clusters at each redshift snapshot, the masses have to be sorted from the smallest to the largest at each redshift step, $M_i \le M_{i+1}$, and the quantity $\bar{\sigma}_{R_0,i}^*$ is defined as

\begin{equation}
\bar{\sigma}_{R_0,i}^* \equiv \frac{1}{2} \left[ \bar{\sigma}_{R_0}(M_i,z) + \bar{\sigma}_{R_0}(M_{i+1},z) \right].
\end{equation}
In practice, this approach is equivalent to assigning the average cross section of clusters with masses $M_i$ and $M_{i+1}$ to all structures with mass between $M_i$ and $M_{i+1}$. The optical depth presented in Eq. (\ref{eqn:tau}) will be a central quantity in the rest of our study and, as mentioned above, we shall set $R_0 = 7.5$ henceforth.

\begin{figure}[t!]
\begin{center}
  \includegraphics[width=\hsize]{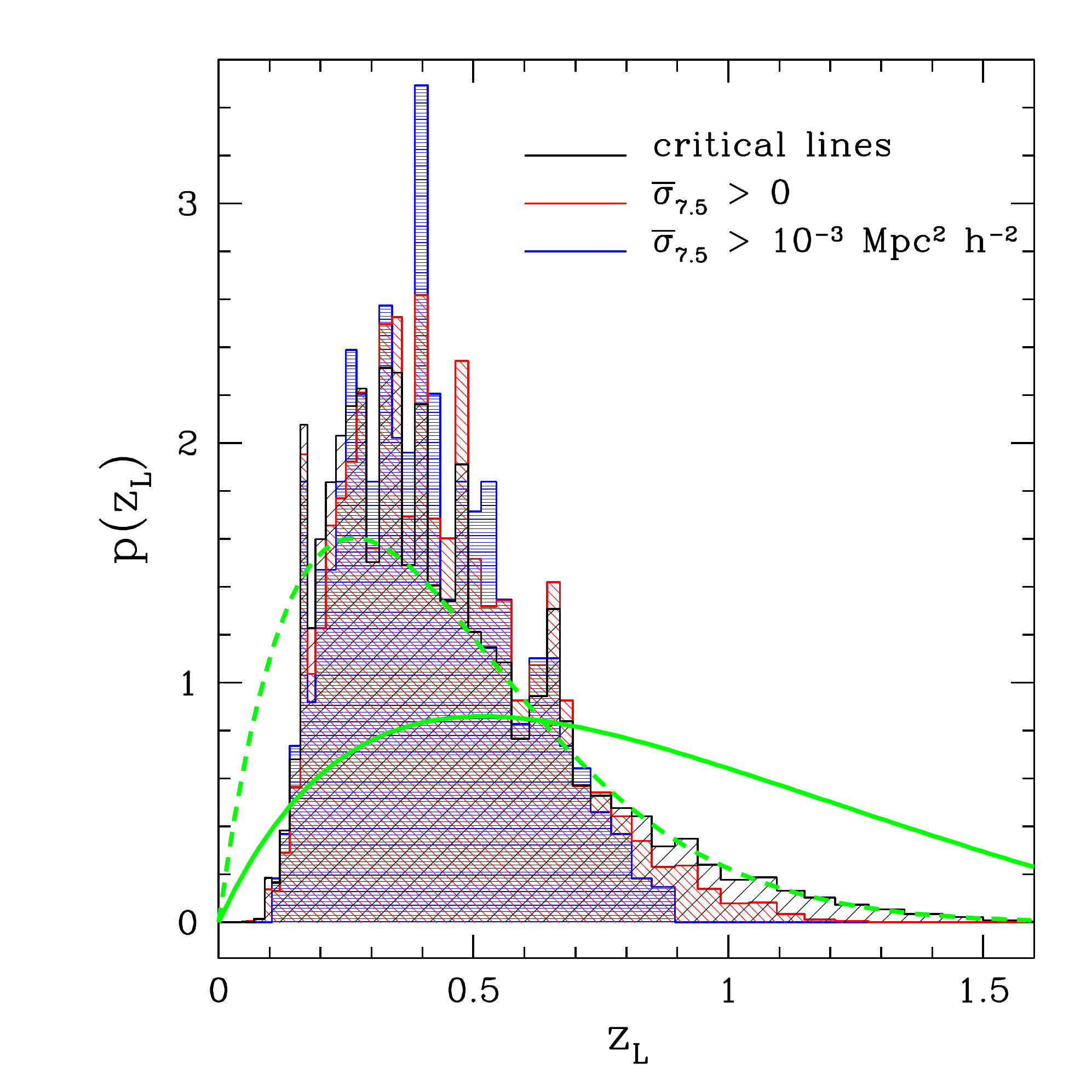}
\end{center}
\caption{The redshift distributions of strong lensing clusters in the {\sc MareNostrum Universe}. The black histogram considers all clusters displaying critical lines, the red one those with a non-vanishing cross section for giant arcs, and the blue one those expected to produce at least one giant arc on average. The thick solid green curve represents the lensing efficiency, given by the normalized lensing distance, while the dashed one represents the lensing efficiency weighted for the cluster abundance (see the text for details).}
\label{fig:zDistribution}
\end{figure}

\section{The population of strong lensing clusters}\label{sct:population}

We begin by reporting the distribution in redshift of the galaxy clusters that are capable of strongly lens background sources in the {\sc MareNostrum Universe}. As done in Paper I we selected three subsamples of such clusters: $i)$ clusters that are capable of producing critical curves, $ii)$ clusters that have a non-vanishing cross section, and $iii)$ clusters that, given suitable background source number counts, are expected to produce at least one giant gravitational arc on average. The cross section of this last class of objects should be around $\sigma_{7.5} \sim 10^{-3}$ Mpc$^2 h^{-2}$. Obviously, the statements above refer to capabilities of at least one projection of the selected object. The strength of gravitational lenses increases when moving from type $i)$ to type $iii)$ since, e.g., the presence of a caustic does not imply a non-vanishing cross section, and the bulk of cross section values is actually lower than $10^{-3}$ Mpc$^2 h^{-2}$ (see Paper I). The redshift distributions for these three categories of lensing clusters are shown in Figure \ref{fig:zDistribution}.

The thick green curve in Figure \ref{fig:zDistribution} represents the lensing efficiency $\varepsilon(z_\mathrm{L})$, that we simply defined as the effective lensing distance normalized such as to have unit integral between $z_\mathrm{L}=0$ and $z_\mathrm{L}=z_\mathrm{S}$. In other words, the lensing efficiency is defined as

\begin{equation}
\varepsilon(z_\mathrm{L}) \equiv \frac{1}{\Upsilon} \frac{D_\mathrm{A}(z_\mathrm{L})D_\mathrm{A}(z_\mathrm{L},z_\mathrm{S})}{D_\mathrm{A}(z_\mathrm{S})},
\end{equation}
where $D_\mathrm{A}(z_1,z_2)$ represents the angular diameter distance between redshifts $z_1$ and $z_2$, and $D_\mathrm{A}(z) = D_\mathrm{A}(0,z)$, while

\begin{equation}
\Upsilon\equiv  \frac{1}{D_\mathrm{A}(z_\mathrm{S})}\int_0^{z_\mathrm{S}} D_\mathrm{A}(z_\mathrm{L})D_\mathrm{A}(z_\mathrm{L},z_\mathrm{S})dz_\mathrm{L}.
\end{equation}
Since the effective lensing distance defines the critical surface density, which is responsible for the formation of critical curves \citep{SU86.1,PA88.1}, it is expected that the redshift trend of $\varepsilon(z_\mathrm{L})$ at least resembles the distribution of strong lensing clusters.

\begin{figure*}[t!]
\begin{center}
  \includegraphics[width=0.7\hsize]{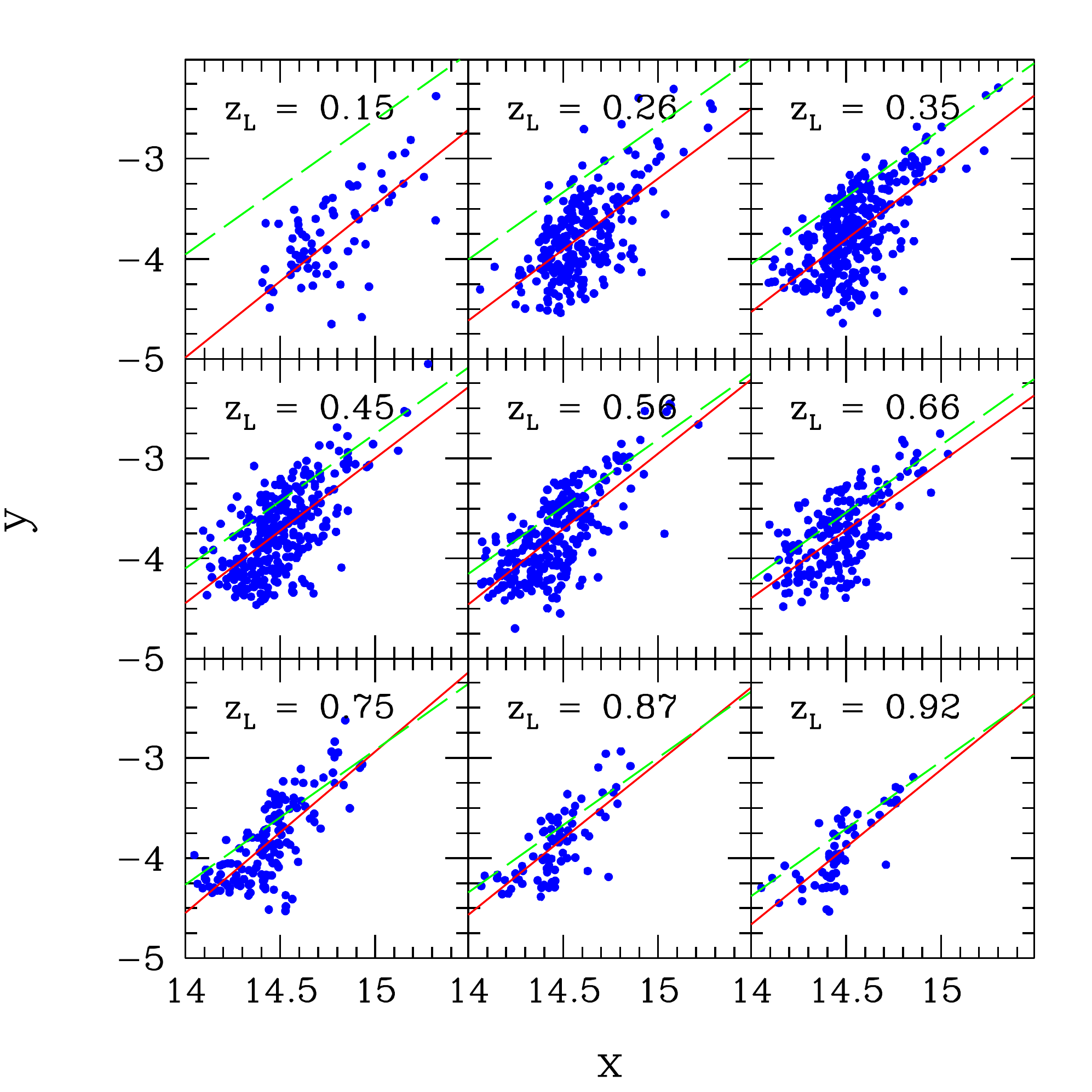}
\end{center}
\caption{The correlation between the logarithm of the virial mass ($x=\log(M\,h/M_\odot)$) and the logarithm of the average cross section for giant arcs ($y=\log(\bar{\sigma}_{7.5}\,h^2/\mathrm{Mpc}^2)$) at several different redshift snapshots of the {\sc MareNostrum Universe}. The red lines show the best linear fit, while the dashed green lines are the expected trend for a SIS lens model (see text for details).}
\label{fig:msigma}
\end{figure*}

This is actually the case, with $\varepsilon (z_\mathrm{L})$ dropping to zero both at $z_\mathrm{L} \simeq 0$ and at $z_\mathrm{L}\simeq z_\mathrm{S}$, as do the simulated distributions. However the redshift distributions of strong lensing clusters are substantially more peaked than the lensing efficiency defined above, and the latter displays a substantial high-redshift tail that is absent in the former. This fact can be understood recalling that the lensing efficiency does not contain any information about the evolution of the cluster mass function. Massive clusters are progressively rarer with increasing redshift, implying that the redshift distribution of strong lensing clusters should also experience an extra drop as compared to that due to pure geometric suppression. This can be verified by additionally weighting the lensing efficiency defined above with the number density of objects that at a given redshift $z_\mathrm{L}$ have mass larger than some fixed threshold value. This is shown by the dashed green curve in Figure \ref{fig:zDistribution}, where we adopted $10^{14} M_\odot h^{-1}$ as the lower mass limit. This roughly corresponds to the minimum mass of strong lensing clusters in the redshift range $0.2 \lesssim z_\mathrm{L}\lesssim 1$. With this extra weighting the high-redshift tail of the simulated distribution of strong lensing clusters is correctly captured, as expected. There remain a disagreement at very low redshifts, due to the fact that very close to the observer the minimum mass capable of producing strong lensing grows rapidly well above $10^{14} M_\odot h^{-1}$. 

Next, let us examine the differences between the redshift distributions of the three categories of strong lensing clusters that we have defined. Going from subsample $i)$ to $iii)$, that is increasing the strength of lenses, we see that the distribution shrinks, with the high-redshift tail being progressively removed and the height of the intermediate-redshift bins increasing. This is an obvious consequence of the fact that the more efficient cluster lenses are expected to be found far away both from the observer and the source plane. As a consequence, no clusters belonging to sample $iii)$ are found beyond $z_\mathrm{L}\sim 0.9$, while clusters merely producing critical lines extend up to $z_\mathrm{L}\sim 1.4$.

\section{Results}\label{sct:results}

\subsection{Lensing efficiency scaling relations}\label{sct:msigma}

\subsubsection{Mass}

As a next point we studied the correlation between the virial mass of galaxy clusters in the {\sc MareNostrum Universe} and their strong lensing efficiency, represented by the cross section for giant arcs. As done before, since for each model cluster we computed the cross section along three different (orthogonal) projections, we quantify the lensing efficiency of each individual cluster as the average $\bar{\sigma}_{7.5}$ over the three projections. In order to simplify the notation, we introduce two new variables, $x\equiv \log (M\,h/M_\odot)$ and $y\equiv \log (\bar{\sigma}_{7.5}\,h^2/\mathrm{Mpc}^2)$.

\begin{figure*}[t!]
\begin{center}
  \includegraphics[width=0.7\hsize]{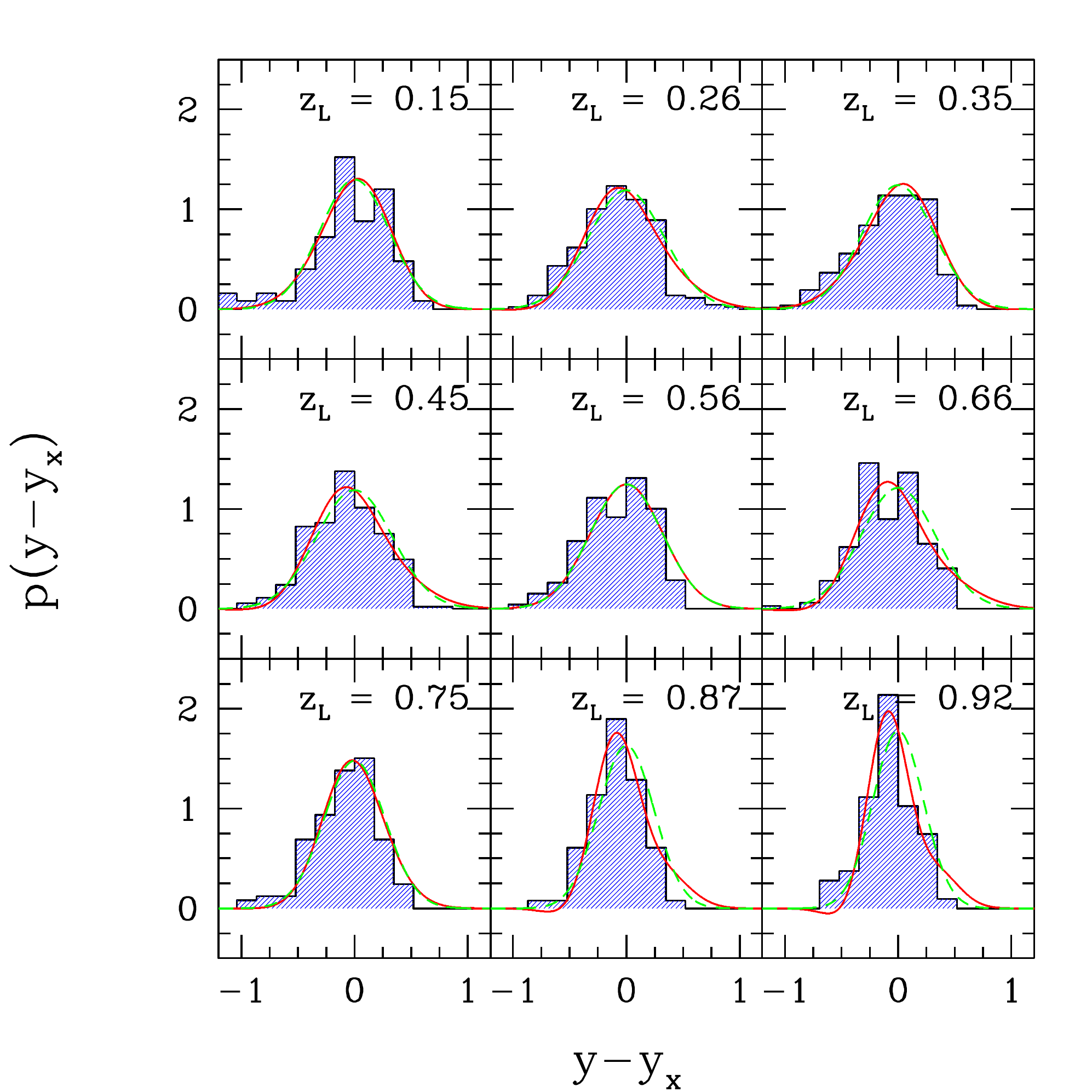}
\end{center}
\caption{The distribution of the logarithm of cross sections for giant arcs ($y=\log(\bar{\sigma}_{7.5}\,h^2/\mathrm{Mpc}^2)$) around the best fit at fixed logarithmic mass ($x=\log(M\,h/M_\odot)$). The green dashed curves show the best Gaussian fit, while the red solid curves show the best first-order correction to the Gaussian fit (see the text for details).}
\label{fig:scatter}
\end{figure*}

In Figure \ref{fig:msigma} we show the correlation between the variables $x$ and $y$ for the clusters of the {\sc MareNostrum Universe} that are contained within various redshift snapshots. For each panel we also show the best linear fit, in the form $y=a(z_\mathrm{L})x+b(z_\mathrm{L})$. We also tried a higher order fit to the points in Figure \ref{fig:msigma}, but the result is not significantly different. We explicitly included the lens redshift dependence of the two best fit parameters $a$ and $b$, since in general their value will differ for different simulation snapshots. The redshift evolution of the best fit parameters is discussed in Section \ref{sct:tau}. As expected, an obvious correlation is present, with more massive objects having on average the larger cross sections. The scatter on the correlation is however considerably large, being up to one order of magnitude both in mass and cross section. This scatter is due to the combination of many different effects, including cluster ellipticity, substructures, and alignment with the line of sight. We investigate more on this scatter further below.

In Figure \ref{fig:msigma} we also reported the expected trend with mass and redshift of the cross section for giant arcs produced by a Singular Isothermal Sphere (SIS henceforth) lens model acting on circular, point-like sources (see \citealt{NA99.1} for a thorough description of this simple lens model). For arcs with length-to-width ratio larger than $R_0$, this cross section can be written as

\begin{equation}
\sigma_{R_0} = \frac{32\pi^3D_\mathrm{A}^2(z_\mathrm{L},z_\mathrm{S})}{c^4}\frac{R_0^2+1}{(R_0^2-1)^2}\sigma_\mathrm{DM}^4,
\end{equation}
where $\sigma_\mathrm{DM}$ is the velocity dispersion of the SIS density profile, that we identify with the average velocity dispersion of dark matter particles within the cluster. We adopted the scaling relation of \citet{EV08.1} in order to relate $\sigma_\mathrm{DM}$ to the mass of clusters, obtaining that

\begin{figure*}[t!]
\begin{center}
  \includegraphics[width=0.7\hsize]{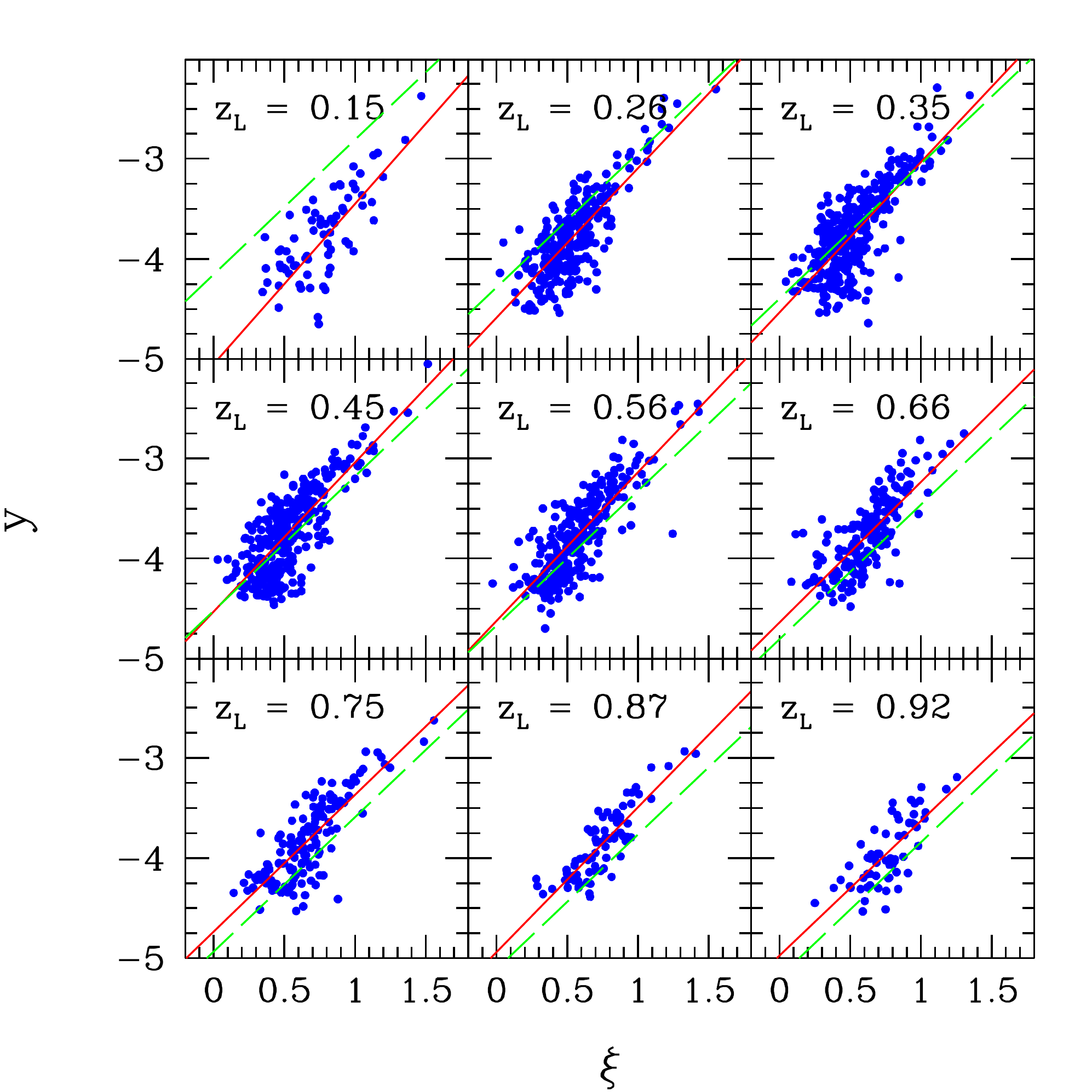}
\end{center}
	\caption{The correlation between the logarithm of the cross section for giant arcs ($y=\log(\bar{\sigma}_{7.5}\,h^2/\mathrm{Mpc}^2)$) and the logarithm of the bolometric X-ray luminosity (in units of $10^{44}$ erg s$^{-1}$, $\xi = \log[L_\mathrm{X}\,\mathrm{s}/(10^{44}\mathrm{erg})]$) for various redshift snapshots of the {\sc MareNostrum Universe} simulation. The red lines show the best linear fits, while the green dashed lines show the self-similar expectation detiled in the text.}
\label{fig:lsigma}
\end{figure*}

\begin{equation}\label{eqn:msigma}
\sigma_{R_0} = \frac{32\pi^3D_\mathrm{A}^2(z_\mathrm{L},z_\mathrm{S})}{c^4}\frac{R_0^2+1}{(R_0^2-1)^2}\sigma_{\mathrm{DM},0}^4\left[ \frac{h(z_\mathrm{L})M}{10^{15}M_\odot} \right]^{4\alpha}.
\end{equation}
In the previous equation we used $\alpha = 0.336$ and $\sigma_{\mathrm{DM},0} = 1.083\times 10^3$ km s$^{-1}$, while $h(z_\mathrm{L}) \equiv H(z_\mathrm{L})/H_0$. The amplitude of the cross sections for the SIS models is almost always larger than the best fit we have found, and this is particularly evident at low redshift. Had we adopted elliptical extended sources for the SIS lenses as well, the relative cross sections would have been increased further, thus worsening the disagreement. In part this can be explained by the fact that what is reported in Figure \ref{fig:msigma} is the average of the lensing cross sections over three orthogonal projections for the {\sc MareNostrum Universe} clusters. Since it is very unlikely that all the three cross sections for each cluster are different from zero, this would downweight the average cross section with respect to the case in which the three cross sections have similar magnitudes, as is the case for SIS lenses (in which obviously $\sigma_{R_0}$ is the same for all projections). Even this however cannot remove the whole disagreement, especially at low redshift and especially if we consider source ellipticity and finite size. Another and more physical effect is that, due to the unrealistically steep density profile, the SIS cross section at low redshifts is expected to be always larger than the cross section of numerically simulated clusters, that tend instead to follow more closely an internally flatter (\citealt{NA96.1}, NFW henceforth) density profile. This fact was already noted by \cite{ME03.1}, where it was also shown that at high redshifts the steepness of SIS lenses compensates quite well the inadequacy of spherical density profiles to reproduce the lensing properties of realistic halos, giving rise to cross sections that are in good agreement with those of the latter. This fact also is confirmed in our Figure \ref{fig:msigma}.

Despite the different normalizations, the slopes of the best fit $x-y$ relations and those derived for the SIS models are intriguingly similar (despite the latter being slightly flatter), especially in light of the large scatter displayed in Figure \ref{fig:msigma}. This implies the interesting fact that, notwithstanding the spread that is introduced in the cluster population by differences in individual formation histories (mergers, substructures, ellipticity, etc.), the simple "self-similar" scaling $\sigma_{R_0} \propto R_\mathrm{E}^2 \propto \sigma_\mathrm{DM}^4 \propto M^{4/3}$ (where $R_\mathrm{E}$ is the Einstein radius of the lens) can be considered a fair representation of average strong lensing clusters.

As noted above the two parameters of the linear fit change with the lens redshift, however we did not find any discernible trend, with both the $a(z_\mathrm{L})$ and $b(z_\mathrm{L})$ functions being rather flat and oscillating around almost fixed values. An exception to this is given by the very low or very high-redshift snapshots, where the fluctuations become very large. However in those snapshots very few cluster with non-vanishing cross sections are available, hence the resulting fit is likely not to be trusted. We come back to this issue again in Section \ref{sct:tau}.

Since the scatter around the best fit relations in Figure \ref{fig:msigma} is substantial, it is necessary to quantify it. Hence, for each strong lensing cluster present in each redshift snapshot we computed the quantity $y-y_x$, where $y_x$ is the logarithm of the cross section given by the best fit relation at that redshift and for a fixed logarithmic mass, while $y$ is the logarithm of the actual cross section. In Figure \ref{fig:scatter} we show how, for each of the redshift snapshots shown in Figure \ref{fig:msigma}, these quantities are distributed.
\begin{figure*}[t!]
\begin{center}
  \includegraphics[width=0.7\hsize]{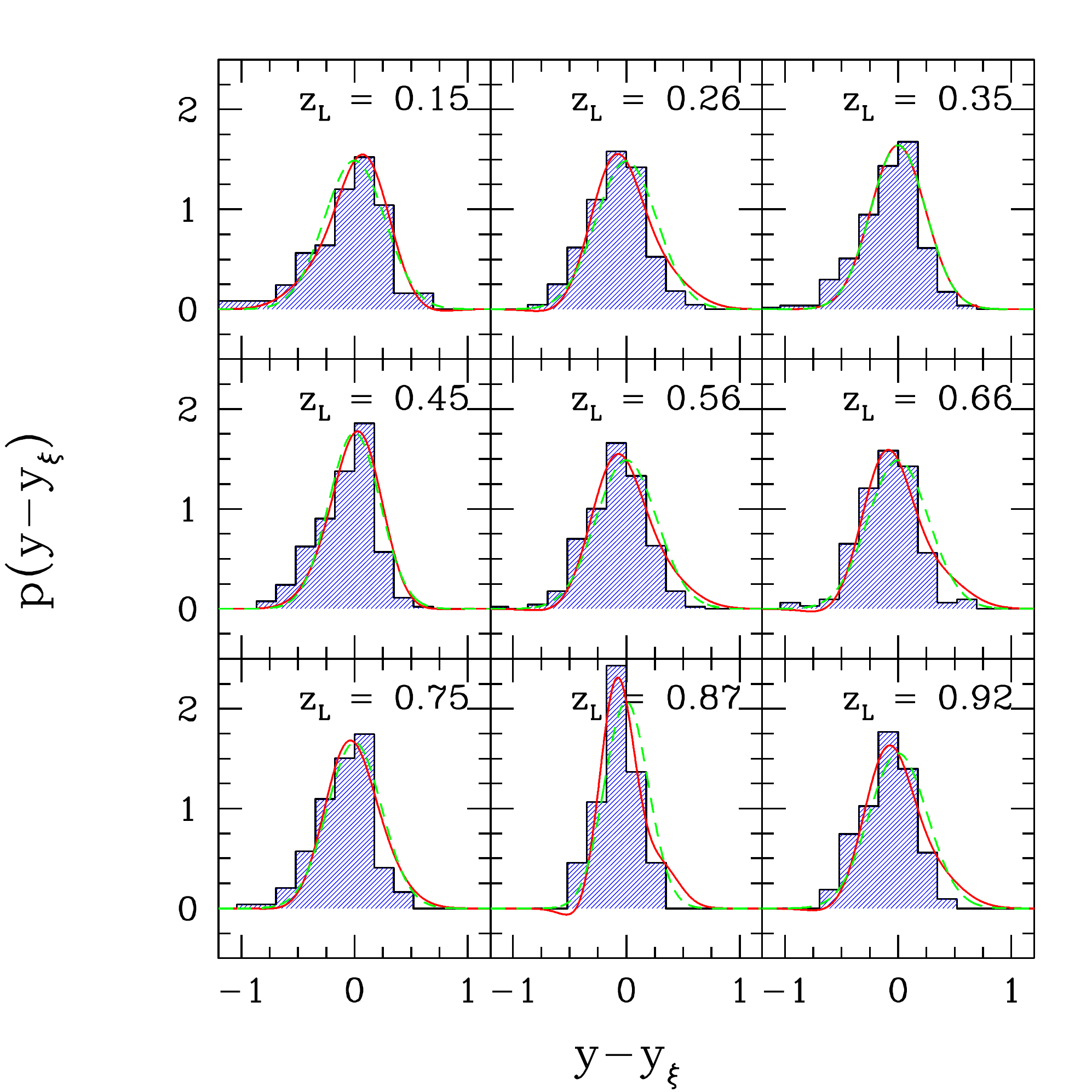}
\end{center}
\caption{The distribution of the logarithm of cross sections for giant arcs ($y=\log(\bar{\sigma}_{7.5}\,h^2/\mathrm{Mpc}^2)$) around the best fit relation at a given X-ray luminosity ($\xi = \log[L_\mathrm{X}\,\mathrm{s}/(10^{44}\mathrm{erg})]$). The green dashed line represents the best Gaussian fit, while the solid red lines are the best non-Gaussian corrections.}
\label{fig:scatter_L}
\end{figure*}
We attempted at first to fit these distributions with Gaussians, with the resulting best fitting curves shown by green dashed curves in Figure \ref{fig:scatter}. The fit is fair but not exceptional, given the rather irregular aspect of the distributions, especially for the high and low-redshift snapshots. Therefore, we added one free parameter representing a suitably small deviation from a Gaussian distribution. In order for this deviation to be related to the moments of the distribution, we adopted the functional form suggested by the Edgeworth expansion \citep{BL98.2,LO08.1}, namely

\begin{equation}\label{eqn:nonGaussian}
p(y-y_x) = \frac{1}{\sqrt{2\pi}\sigma}\exp\left[-\frac{(y-y_x)^2}{2\sigma^2}\right] \left[ 1+\sigma H_3(y-y_0)\frac{S_3}{6} \right],
\end{equation}
where $H_3(u)$ is the Hermite polynomial of third degree, that reads

\begin{equation}
H_3(u) = u^3-3u.
\end{equation}
The parameter $S_3 = S_3(z_\mathrm{L})$ is the normalized skewness of the distribution, and it has been left as the second free parameter. Obviously if $S_3 = 0$ we fold back to the Gaussian case where the only free parameter is $\sigma = \sigma(z_\mathrm{L})$ (not to be confused with the lensing cross section).

In Figure \ref{fig:scatter} the red solid curves show the results of this kind of non-Gaussian fit. As can be seen, Eq. (\ref{eqn:nonGaussian}) provides a slight improvement over the simply Gaussian fit, especially for low and high-redshift snapshots. It is likely that the fit could be improved further by introducing a third free parameter (that in the Edgeworth expansion would represent the normalized kurtosis). Also, there are several snapshots (e.g., those at $z=0.56$ and $z=0.66$) that are showing traces of bimodality, for which a fit with the superposition of two curves would be preferable. As we discuss in Section \ref{sct:tau} the two parameter fits that are shown in Figure \ref{fig:scatter} are not precise enough to accurately reproduce the strong lensing optical depth measured in the {\sc MareNostrum Simulation}. Since however we have found that the inclusion of the scatter is not necessary for this scope, we prefer to add no further complication and stick to the two parameter fits. Besides, while the redshift evolution of $\sigma(z_\mathrm{L})$ is rather flat around an average of $\sim 0.3$, the redshift evolution of the parameter $S_3(z_\mathrm{L})$ is much more noisy than for $a(z_\mathrm{L})$ and $b(z_\mathrm{L})$, showing large fluctuations between neighboring redshift snapshots. This fact suggests that with the introduction of the skewness as an additional fit parameter we are in part overfitting the noise of measured distributions. The addition of further parameters would worsen this situation even more, thus being of little practical use.

\subsubsection{Luminosity}

Since the {\sc MareNostrum Universe} simulation includes adiabatic gas, we could consider the correlations of the virial mass and the strong lensing cross section with the bolometric X-ray luminosity, the latter being a quantity that is more directly related to observable features. To that end, we introduced a new variable $\xi \equiv \log\left[L_\mathrm{X}\;\mathrm{s}/\left(10^{44}\mathrm{erg}\right)\right]$, where $L_\mathrm{X}$ is the bolometric X-ray luminosity. In Figure \ref{fig:lsigma} we report the relation between $y$ and $\xi$, while the $\xi-x$ relation has been already discussed in Paper I. 

In the same Figure we also report the best linear fit (red solid lines) and the $y-\xi$ relation that can be deduced by combining the SIS scaling derived above for the $y-x$ relation with self-similar arguments (green dashed lines), according to the following line of reasoning. Let us assume that dark matter halos in the {\sc MareNostrum Universe} simulation follow on average a NFW density profile. Self-similar scaling arguments imply that the bolometric luminosity of a plasma in hydrostatic equilibrium in the potential well of a dark matter halo with such a density profile follows \citep{EK98.1,FE07.2}

\begin{equation}
L_\mathrm{X}(M,z_\mathrm{L}) \propto M^{4/3}\frac{c^{7/2}}{F(c)^{5/2}} h(z_\mathrm{L})^{7/3},
\end{equation}
where $z_\mathrm{L}$ is the redshift of the lensing cluster at hand, $M$ is its virial mass, and $c$ is its concentration. The function $F(c)$ reads

\begin{equation}
F(c) = \ln(1+c) - \frac{c}{1+c}.
\end{equation}
Since even for the most massive clusters the concentration is at least of a few, we have that $F(c) \simeq \ln(c)-1$, and this dependence, being very weak, can be neglected. We further recall that, roughly, $c \propto M^{-0.1}$ thus, neglecting the redshift dependence of the concentration, overall we obtain the approximate scaling $L_\mathrm{X}(M,z_\mathrm{L}) \propto M\,h(z_\mathrm{L})^{7/3}$. In combination with Eq. (\ref{eqn:msigma}) this finally implies

\begin{equation}
\sigma_{R_0} \propto L_\mathrm{X}^{4\alpha} \frac{D_\mathrm{A}^2(z_\mathrm{L},z_\mathrm{S})}{h(z_\mathrm{L})^{16\alpha/3}}.
\end{equation}
In order to fix the normalization we simply required that the bolometric X-ray luminosity and cross sections of the most luminous cluster in the simulation are reproduced. We would like to stress that the procedure we followed for deriving of the above scaling relations is likely oversimplified. Here we did not want to perform a completely rigorous derivation, just to make the case for the fact that simple scaling arguments can produce correlations that are rather similar in slope to those observed in the {\sc MareNostrum Universe}.

Because of the increase in average X-ray luminosity with increasing mass and because of the trend reported in Figure \ref{fig:msigma}, it is expected that the lensing cross section also increases with the X-ray emission. This is indeed observed in Figure \ref{fig:lsigma} and the correlation is visibly tighter than the $y-x$ correlation. In order to better appreciate this, in Figure \ref{fig:scatter_L} we report the distribution of logarithmic cross sections around the best fit relation for a fixed X-ray luminosity. As a matter of fact, the logarithmic slope of the $y-\xi$ relation is relatively similar to that of the $y-x$ relation (in agreement with the self-similar scaling detailed above), but the scatter around the latter is systematically $\sim 25\%$ smaller than around the former. Also, as we already found for the mass-cross section relation, the introduction of a non-Gaussian part to the functional form used for fitting the distribution around the best suitable luminosity-cross section relation does not substantially alter the fit with respect to the perfectly Gaussian form.

The fact that the scatter around the best fit relation is decreased suggests that there should be a more fundamental relation of the cross section for giant arcs with the bolometric X-ray luminosity, rather than with the virial mass. This can be somewhat expected from several qualitative considerations: a) relaxed clusters are expected to have large gas and dark matter central densities. This would increase both the X-ray emissivity, that is proportional to the square of the gas density, and the extension of the critical curves \citep{FE07.2}; b) cluster mergers are expected to compress and heat up the intracluster medium, hence increasing the X-ray luminosity. At the same time, interactions should increase the convergence and shear fields of the cluster \citep{TO04.1}, thus enhancing its lensing efficiency; c) in general, the X-ray luminosity is particularly sensitive to the physical processes taking place in the inner region of galaxy clusters, and so is the strong lensing efficiency. The virial mass on the other hand is much less sensitive to that. This also implies that the correlation between the lensing cross section and the mass contained within smaller radii (e.g., $M_{500}, M_{2500}$, etc.), should be tighter than with the virial mass.

A closer inspection of Figure \ref{fig:lsigma} shows that virtually all the clusters expected to have at least one giant arc ($\bar{\sigma}_{7.5} \gtrsim 10^{-3}$ Mpc$^2 h^{-2}$) have a bolometric X-ray  luminosity that is larger than $\sim 6\times 10^{44}$ erg s$^{-1}$. The converse is in general not true, in the sense that many clusters exist with a high luminosity that make rather poor strong lenses. Hence, strong lensing clusters would be good tracers of centrally dense intra-cluster medium, while the converse would not hold in general. 

Finally, we tried to recompute the scaling relations between $x$ and $\xi$ and the logarithm of the cross section $y$ by taking into account only relaxed clusters. Relaxation was quantified through the parameter $\beta$ introduced in Paper I, that expresses deviations from virial equilibrium: negative $\beta$ implies systems dominated by the kinetic energy, positive $\beta$ by the potential energy, while $\beta=0$ implies perfect virial equilibrium. We recomputed the above scaling relations by using only clusters with $\left|\beta\right|\le0.3$, and found no reduction in the scatter about the best fit $x-y$ and $\xi-y$ relations, nor a significant shift of the quartiles of the distributions. What happen instead is just a removal of the highest mass/luminosity objects, which is expected since those are the least relaxed on average. This is an indication that, intriguingly, the bulk of this scatter should not be originated by the dynamical activity of clusters, rather by triaxiality and concentration distributions. The latter, particularly,
is expected to have a major impact since the cross section is remarkably sensitive to it, and the concentration itself has a substantial intrinsic scatter. Because of the relatively low resolution of clusters in the simulation, we could compute only concentrations for staked cluster samples, not for individual objects. Hence we could not asses this issue in detail.

\subsection{Optical depths}\label{sct:tau} 

\subsubsection{Synthetic optical depth}

\begin{figure*}[t!]
\begin{center}
  \includegraphics[width=0.45\hsize]{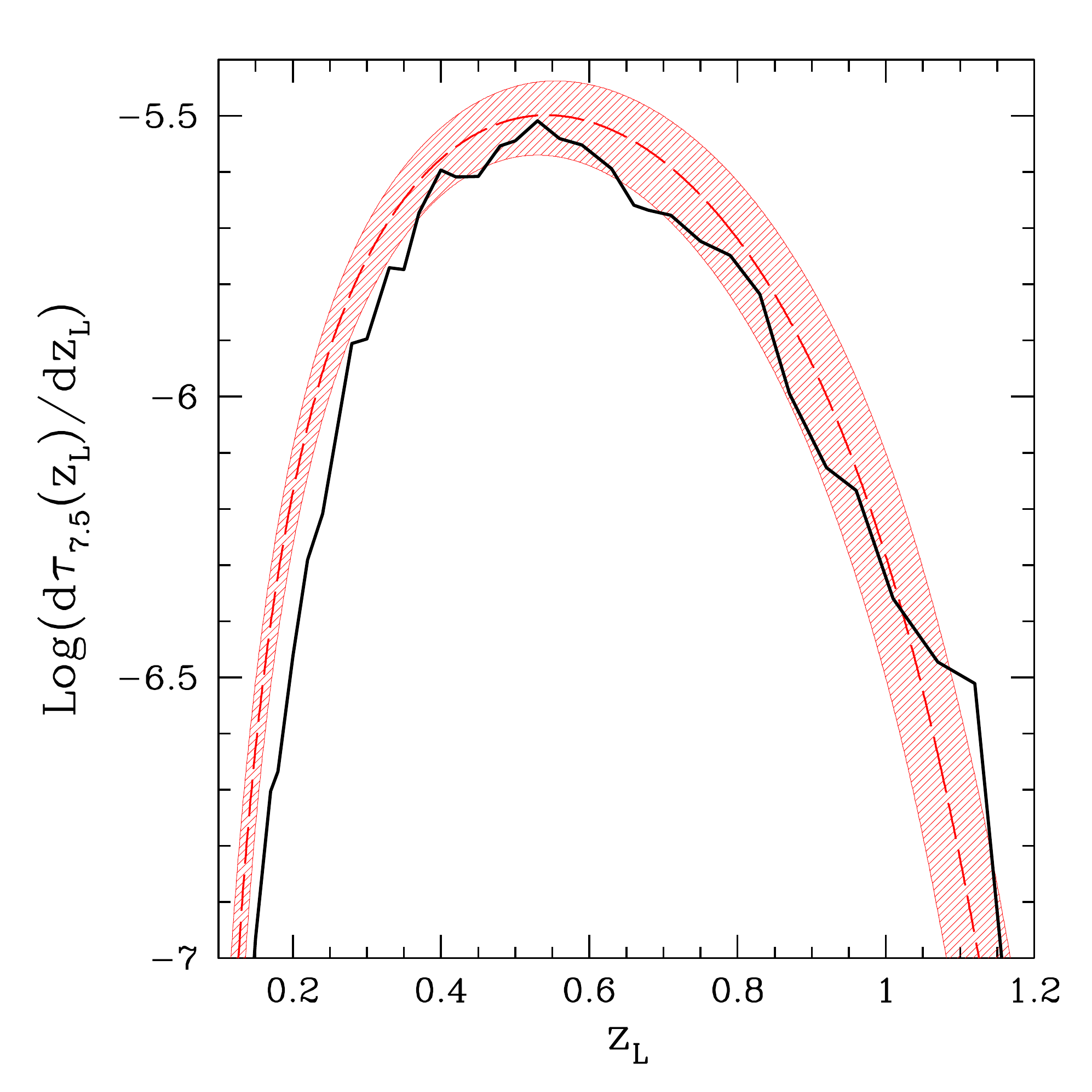}
  \includegraphics[width=0.45\hsize]{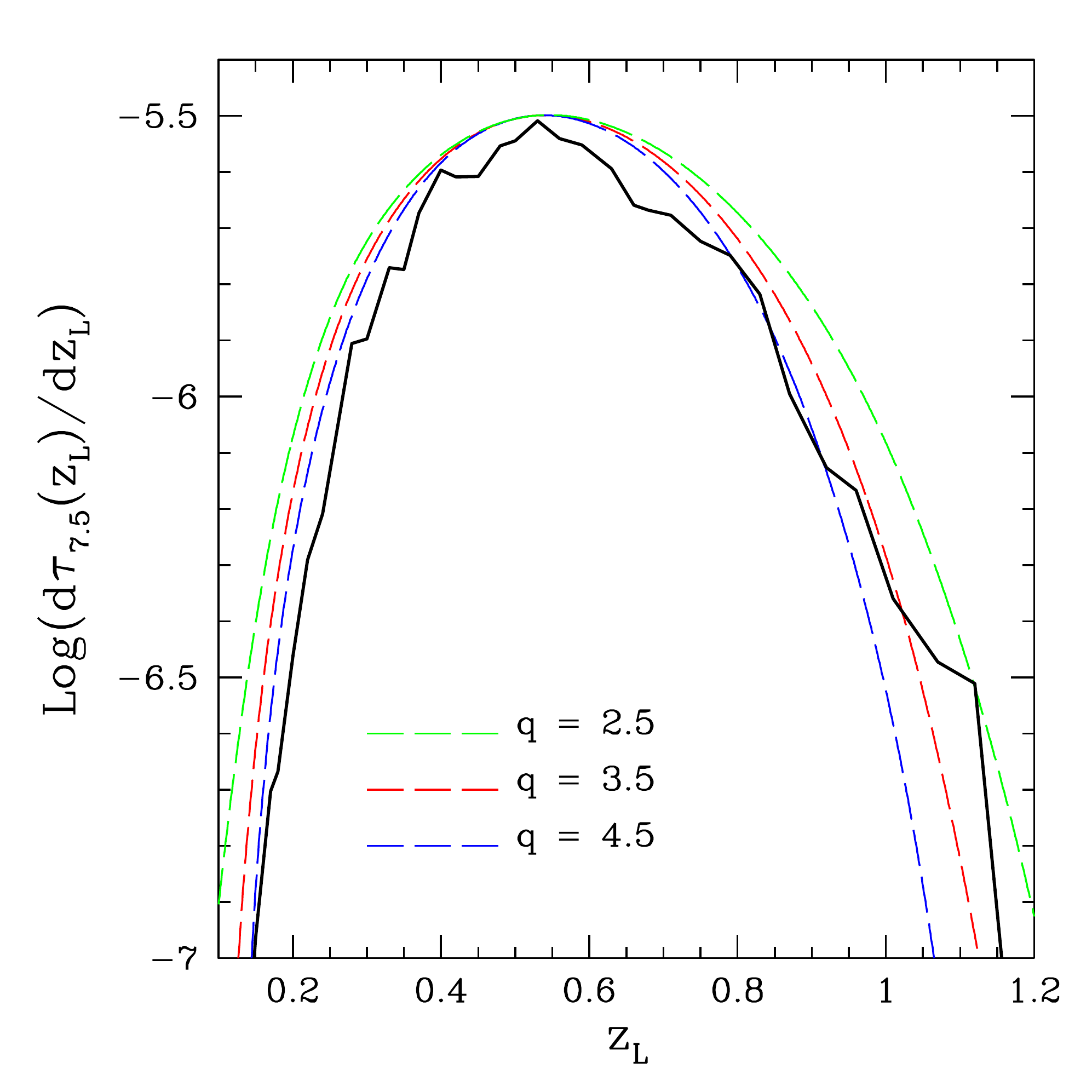}\hfill
\end{center}
	\caption{\emph{Left panel}. The differential optical depth for giant arcs obtained by all clusters in the {\sc MareNostrum Universe} as a function of lens redshift (black solid line). The red dashed line represents the 	synthetic differential optical depth obtained with the procedure outlined in the text ($q=3.5$), while the shaded area represents the effect of a $30\%$ uncertainty on the minimum mass at the pivotal redshift (see the text for details). \emph{Right panel}. Same as the left panel but with the blue and green dashed lines now representing the synthetic optical depths computed with values of $q$ different from $3.5$, as labeled.}
\label{fig:tau}
\end{figure*}

In Figure \ref{fig:tau} we show the differential optical depth as a function of lens redshift. The differential optical depth is the integrand of Eq. (\ref{eqn:numtau}), and represents the contribution to the total optical depth that is given by structures at different lens redshifts. As a consequence, the integral under the black solid curve in Figure \ref{fig:tau} represents the total optical depth, that for sources at $z_\mathrm{S} = 2$ amount to $\tau_{7.5}(z_\mathrm{S} = 2) \simeq 1.59\times 10^{-6}$. The differential optical depth has the generic shape that is expected, namely dropping at low and high redshift due to the geometric suppression of the lensing efficiency. The total optical depth is substantially lower than that reported for a $\sigma_8=0.9$ cosmology in \citet{FE08.1}, which is to be expected since there the authors used a complete source redshift distribution, while here we adopted sources at a fixed redshift. By multiplying the total optical depth with the number counts of source galaxies at $z_\mathrm{S} = 2$ it is possible to infer the abundance of arcs produced by those sources that are visible below some given magnitude threshold (see the discussion is Section \ref{sct:simulation}). 

The existence of a relation between the virial mass of galaxy clusters and their cross sections for giant arcs (Section \ref{sct:msigma}) implies that we could in principle construct a synthetic realization of the optical depth in the following way. For each step in redshift we randomly extract a large number of masses, associate a cross section to each mass through the $x-y$ relation described above, suitably randomized according to the scatter around the best fit relation, and use these cross sections for computing the differential optical depth at the fixed redshift. This kind of procedure would allow one to compute strong lensing optical depths based only on the scaling relations that we have found and without expensive calculations involving ray-tracing simulations and cross section evaluations. 

\begin{figure}[t!]
\begin{center}
  \includegraphics[width=\hsize]{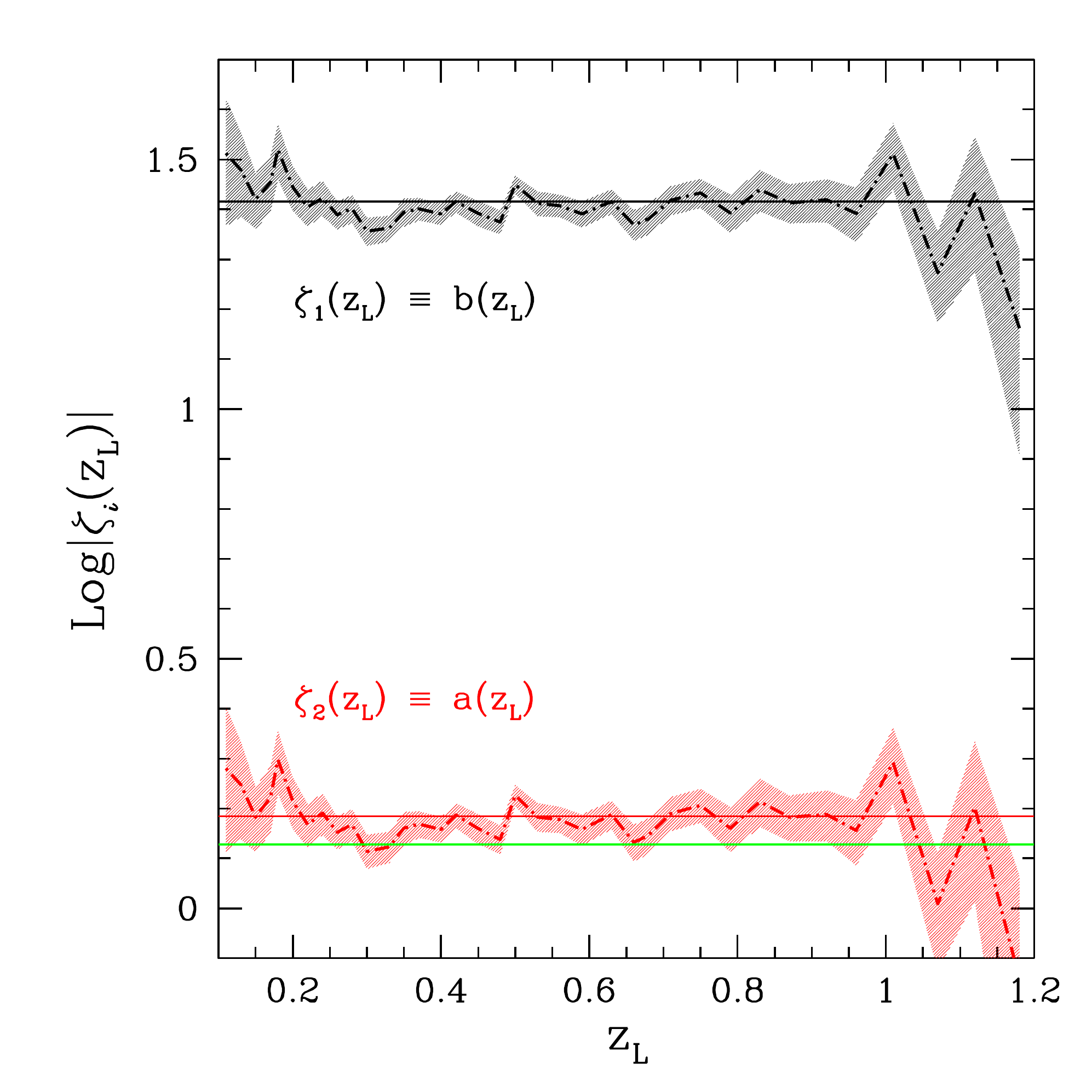}
\end{center}
\caption{The redshift evolution of the two parameters defining the best fit $x-y$ relation (see the text for details), with the respective standard errors. The black and red solid horizontal lines represent the respective redshift averages, while the green horizontal line represents the slope expected for a SIS lens model.}
\label{fig:bestFit}
\end{figure}

In Figure \ref{fig:bestFit} we show the redshift evolution of the parameters $a(z_\mathrm{L})$ and $b(z_\mathrm{L})$ defined in the previous Section \ref{sct:msigma}. As announced, both of them oscillate around a roughly constant value. We plot in the same Figure also the average values, computed over the redshift range $[0.1,1.1]$, since outside this range very little lensing clusters are present per redshift snapshot, and the result of the fitting procedure is less reliable. These two average values amount to $\bar{a} = 1.529$ and $\bar{b} = -26.08$. Although the procedure described above would make use of the specific vales $a(z_\mathrm{L})$ and $b(z_\mathrm{L})$ at each redshift $z_\mathrm{L}$, this would not be of very practical use. Therefore we decided to adopt the same values of the two parameters $\bar{a}$ and $\bar{b}$ at each redshift snapshot, equal to the respective average values. At first, we tried to ignore the scatter around the best fit $y-x$ relation, i.e., we assigned at each mass the exact value that is prescribed by the linear logarithmic fit with the redshift averaged values. Another factor that needs to be taken into account is the smallest mass $M_\mathrm{min}(z_\mathrm{L})$ that, at each redshift step, enters in the computation of the differential optical depth ($M_1$ in Eq. \ref{eqn:numtau}). As it turns out this point is of fundamental importance, since the optical depth tends to be dominated by the smallest clusters capable of producing a non-vanishing cross section, being by far the more abundant. Instead of adopting as such lower limit the smallest mass that at each redshift produces a non-vanishing cross section in the {\sc MareNostrum Universe} we let the redshift evolution of $M_\mathrm{min}(z_\mathrm{L})$ be

\begin{equation}\label{eqn:mmin}
M_\mathrm{min}(z_\mathrm{L}) = M_\mathrm{min, p} \left[\frac{\varepsilon(z_\mathrm{L,p})}{\varepsilon(z_\mathrm{L})}\right]^q.
\end{equation}
We adopted this procedure for a twofold reason. Firstly, choosing a different minimum mass at each redshift snapshot is not practical, exactly as it was not practical to use a different pair of best fitting parameters at each redshift. Second, it is likely that this minimum mass is siumlation-dependent, or in different words, that rerunning the same simulation several times would produce a distribution of minimum masses at a given redshift, due to the randomness of the cluster formation histories.

In Eq. (\ref{eqn:mmin}) the function $\varepsilon(z_\mathrm{L})$ is the lensing efficiency that has been introduced in Section \ref{sct:population}, $z_\mathrm{L, p}$ is some pivotal lens redshift and $M_\mathrm{min, p}$ is the corresponding minimum lensing mass. We adopted as pivotal the redshift at which the differential optical depth is maximal, $z_\mathrm{L,p}=0.53$, that also coincides with the broad maximum in the lensing efficiency reported in Figure \ref{fig:zDistribution}. At this redshift the smallest mass with non-vanishing cross section is $M_\mathrm{min, p} = 10^{14}M_\odot h^{-1}$. 
Ideally, selecting a suitable value of $q$ for each lens redshift would allow to match the synthetic optical depth, since one would encapsulate in this free parameter all the remaining uncertainty that has not been taken into account, i.e., the deviation of the parameters $a(z_\mathrm{L})$ and $b(z_\mathrm{L})$ from their redshift averages, the scatter around the best fit $x-y$ relation, etc. In practice, we can choose a single value of $q$ that gives an overall good fit to the true optical depth.

The "synthetic" optical depth resulting from this procedure is represented in Figure \ref{fig:tau}, together with the effect of allowing some uncertainty on the value of $M_\mathrm{min,p}$ and changing $q$ within some reasonable interval. The deviation of the synthetic optical depth from the true one for, e.g.,  $q = 3.5$ might seem quite significant. Particularly, the true optical depth appears to be quite overestimated at $z_\mathrm{L} = 0.2-0.3$ due to the fact that both the parameters $a(z_\mathrm{L})$ and $b(z_\mathrm{L})$ are below the average at those redshifts (see Figure \ref{fig:bestFit}). However it should be recalled that the Figure has logarithmic scale, and as a matter of fact the synthetic optical depth with $q = 3.5$ differs from the true one only by about $\sim 10-15\%$, which is a truly negligible amount given the large uncertainties involved in the estimation of arc abundances. Increasing the value of $q$ to, e.g., $4.5$ reduces the discrepancy on the total optical depth to only $\sim 5\%$, however in this case the high-redshift tail of the differential optical depth is rather underestimated. Based on these numbers and on Figure \ref{fig:tau} we suggest to adopt a value of $q \simeq 3.5$ as a fair compromise.

\begin{figure*}[t!]
\begin{center}
  \includegraphics[width=0.7\hsize]{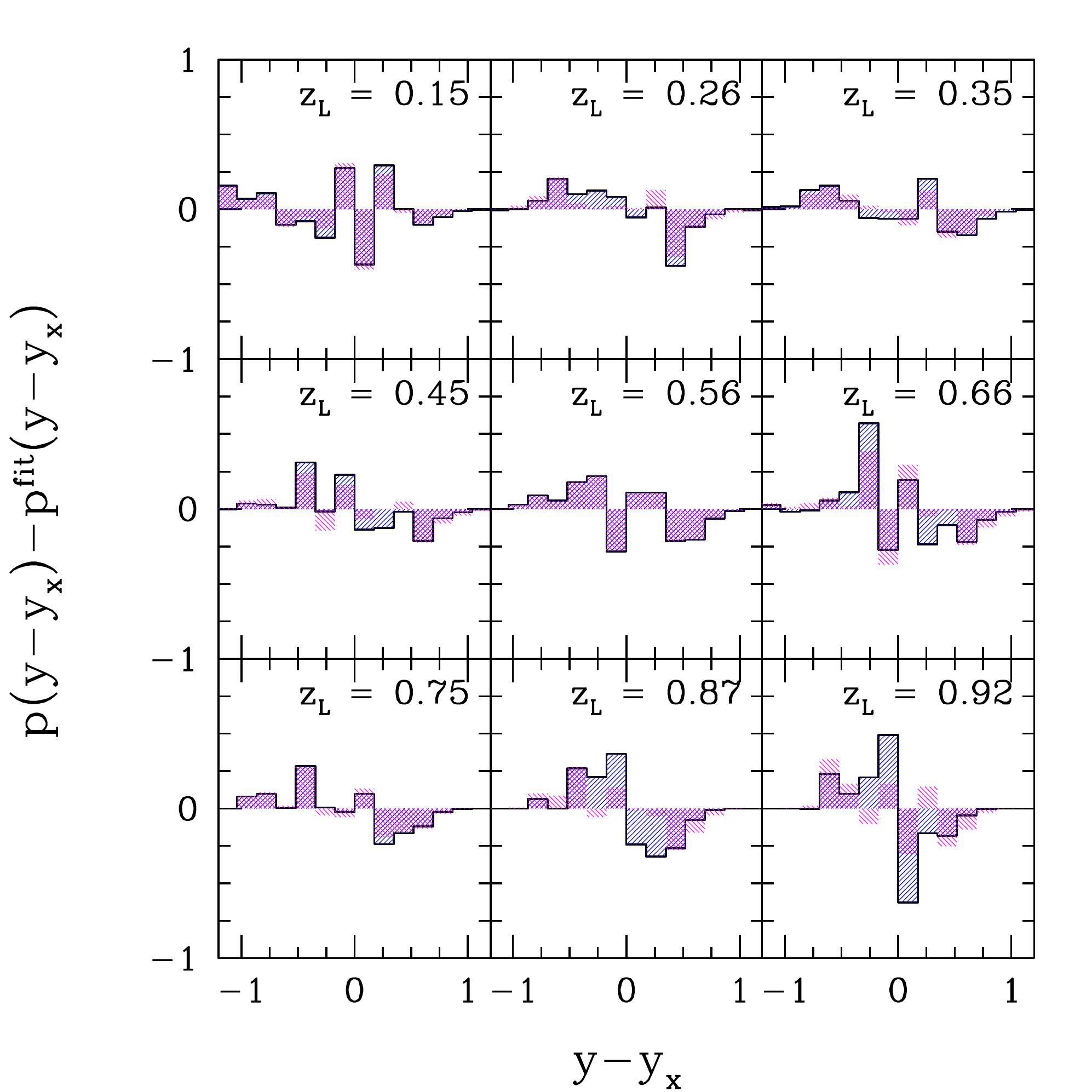}
\end{center}
\caption{The difference between the true distribution around the best fit mass-cross section relation and the best Gaussian (blue shaded histograms with black contours) and non-Gaussian (magenta shaded histograms) fits. Results for several different redshift snapshots are shown, as labeled in the plot.}
\label{fig:difference}
\end{figure*}

Also shown in Figure \ref{fig:tau} is the effect of uncertainty in the value of $M_\mathrm{min, p}$. As a matter of fact, a substantial fluctuation on the minimum lensing mass at redshifts close to the pivotal one is found, due to the fact that in such a boundary situation small variations in the internal structure of the lens are enough to scatter the lens itself above or below the threshold for cross section production. In Figure \ref{fig:tau} we show as an example the effect of a $30\%$ uncertainty on $M_\mathrm{min,p}$, which can easily shift the total optical depth of $\sim 50\%$. The effect of changing the value of $q$ within a reasonable interval is relatively similar in amplitude to the effect of $M_\mathrm{min, p}$ fluctuations. 

We have to observe that the reasonable way in which the synthetic optical depth approximates the true one is heavily dependent on the values of $\bar{a}$ and $\bar{b}$ that we assume. Particularly, to produce variations to the synthetic optical depth with magnitude similar to that reported in the left panel of Figure \ref{fig:tau}, it is sufficient to modify the normalization or the slope of the mass-cross section relation at the percent level. This is way smaller than the fluctuations and errors that are reported in Figure \ref{fig:bestFit}. For instance, adopting the slope of the mass-cross section relation that is found for SIS lens model, instead of the redshift average that we adopted, would produce a gross underestimate of the true optical depth. This argument can be turned around by saying that the synthetic optical depth is a fair representation of the true one only if the normalization and the slope of the mass-cross section relation adopted stay very close to their redshift-average values computed here.

Lastly, we would like to comment on the fact that in the above procedure for computing synthetic optical depth we neglected the role of the scatter around the best fit $y-x$ relation. As we mentioned in Section \ref{sct:msigma}, the fits that we performed to the distributions of $y-y_x$ both with a Gaussian and a non-Gaussian functions are not precise enough for the purpose of increasing the precision of the synthetic optical depth. This can be understood as follows: as it appears evident from a close inspection of Figure \ref{fig:scatter}, neither the Gaussian nor the non-Gaussian functional forms fully appreciate the asymmetry in the scatter of cross sections around the best fit. Namely, the distributions of $y-y_x$ tend to be substantially skewed toward negative values as compared to the best fits. As a consequence, the introduction of the scatter around the best fit actually worsen the agreement of the synthetic optical depth to the true one. This fact can be better appreciated by analyzing Figure \ref{fig:difference}. There, the scatter around the best fit mass-cross section relation is shown once the best Gaussian and non-Gaussian fits were removed. It is obvious that the best Gaussian fit is substantially more positively skewed with respect to the true distribution, especially at high redshift. The same remains true for the non-Gaussian fit, although to a lesser extent. This means that, including the scatter in the computation of the synthetic optical depth effectively equals at assigning higher cross sections to a given mass with respect to reality. As a result the synthetic optical depth tends to be a substantial overestimate of the true one if the scatter is included 

In order to solve this, a better fit to the distributions of $y-y_x$ would be required, maybe with the superposition of two curves. This however would increase the number of free parameters from two to four at least. We chose not to pursue this route, since it would add substantial complication to a model that already works quite acceptably. As a matter of fact, given the very simplistic nature of the underlying assumptions (one best fit relation valid for all redshifts, absence of scatter, etc.) it is very remarkable that the true optical depth can be reproduced at this level with just one adjustable parameter, $q$. This very simple model for computing synthetic optical depths can be very helpful for, e.g., evaluate the contribution to arc abundance that is given by structures in various redshift ranges without performing any actual strong lensing calculation. Summarizing, in evaluating synthetic optical depths for sources at $z_\mathrm{S}=2$ we suggest to stick to the precise values of $\bar{a}$ and $\bar{b}$ that we adopted, and to use a value of $q$ between $3.5$ and $4.5$, with the threshold mass at the pivot redshift equaling $10^{14} M_\odot h^{-1}$.

\subsubsection{The role of dynamical activity}

\begin{figure}[t!]
\begin{center}
  \includegraphics[width=\hsize]{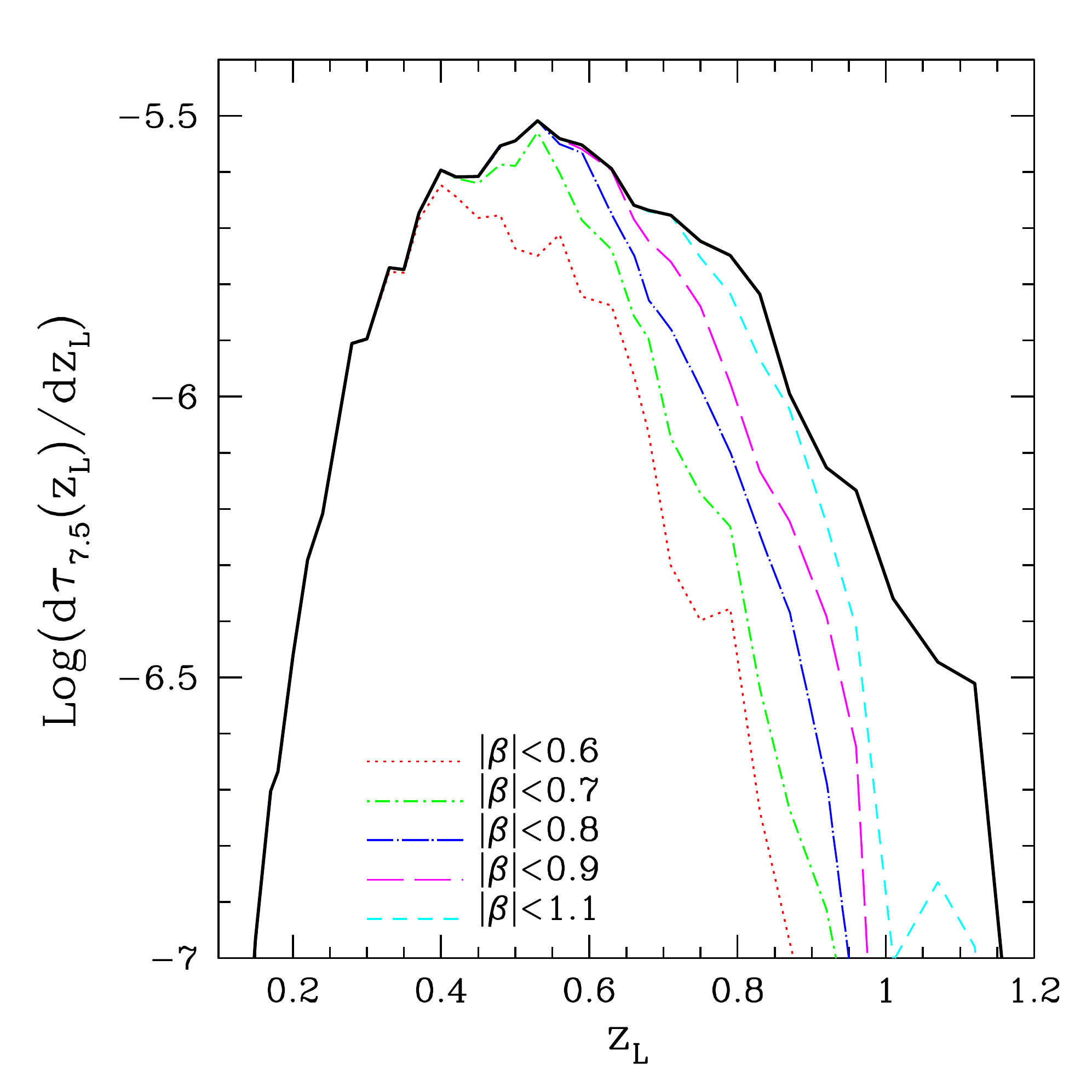}\hfill
\end{center}
	\caption{The differential optical depth in the {\sc MareNostrum Universe} computed for all clusters in the simulation (black solid line) and by including only those clusters whose $\beta$ parameter is included in some neighborhood of zero (dashed colored lines), as labeled in the plot.}
\label{fig:betaTau}
\end{figure}

In Figure \ref{fig:betaTau} we show the differential optical depth of the {\sc MareNostrum Universe} obtained by excising those clusters that are not relaxed enough. The relaxation of a cluster was established via the virial equilibrium parameter, defined in Paper I (see also \citealt{SH06.1}). As can be seen the exclusion of clusters with a value of $\beta$ significantly different from zero causes an erosion of the high-redshift tail of the differential optical depth. Particularly, when only clusters with $|\beta|<0.6$ are included, the differential optical depth peaks at $z_\mathrm{L} \sim 0.4$ instead of $z_\mathrm{L} \sim 0.55$, and the total optical depth is reduced by a factor of 2. In other words, half of the number of giant arcs produced by sources at $z_\mathrm{S} = 2$ are expected to be found inside clusters that are substantially unrelaxed. A significant contribution to arc statistics from clusters at $z_\mathrm{L} > 0.8$ can be expected only by including objects with $|\beta|\lesssim 1$.

This trend of the optical depth with the virial equilibrium of clusters can be easily understood by looking at the redshift evolution of $\beta$ that has been presented in Paper I. Namely, while at $z\sim 0$ the average $\beta$ is quite close to zero, at high redshifts it tends to become strongly negative. This means that, removing clusters with $|\beta|$ very different from zero is equivalent at removing high-redshift clusters, and hence the drop in the differential optical depth at large $z_\mathrm{L}$ follows. Intriguingly, the redshift distribution of strong lensing clusters carries information on the dynamical state of the deflectors: selecting clusters that are capable of producing giant gravitational arcs at $z_\mathrm{L}\lesssim0.5$ we are automatically selecting objects that are relatively close to the virial equilibrium. 
Vice versa, lensing clusters at high redshift are very likely to be dynamically active.

We wish to interpret this result in light of the incidence of gravitational arcs in high-redshift clusters found by, e.g., \citet{GL03.1} (see also \citealt{ZA03.1}). Particularly, \citet{GL03.1} analyzed the Red-sequence Cluster Survey (RCS, \citealt{GL01.1}), finding eight clusters at $z_\mathrm{L} \gtrsim 0.64$ with prominent gravitational arcs, for a total of $11$ strong lensing features, including tentative detections. By using Figure \ref{fig:betaTau} we can estimate that $\sim 77\%$ of the optical depth at $z_\mathrm{L} \gtrsim 0.64$ is produced by clusters with $|\beta| > 0.6$, which effectively means $\beta < -0.6$ since the values of $\beta$ are prominently negative, especially at high redshift (see Paper I). Thus, it is expected that $\sim 8-9$ of the $11$ giant arcs of \cite{GL03.1} are produced by substantially unrelaxed clusters. As a matter of fact, \cite{GL03.1} themselves argued that the high-redshift lensing clusters they found are somewhat enhanced in their lensing efficiency by some physical process, possibly the presence of substructures and dynamical activity. 

Browsing the literature in order to confirm the fraction of strong lensing clusters that are dynamically unrelaxed gives inconclusive results. For instance, RCS 2319.9+0038 is part of a supercluster and is likely to be unrelaxed, as well as RCS 2156.7-0448 \citep{HI08.1}. RCS 1620.2+2929 was shown by \citet{GI07.1} to have an excess velocity dispersion compared to what would be expected by its optical richness \citep{YE03.1}, suggesting signs of dynamical activity. On the other hand, RCS 0224.5-0002 is likely a relaxed object \citep{HI07.1}, totaling to 4 giant arcs found in dynamically active clusters versus 2 found in relaxed clusters. All the other RCS clusters used by \citet{GL03.1} do not have information about the dynamical state. Obviously we are not able to draw any conclusion from this, but just want to stress that further study of the dynamical activity of strong lensing clusters, especially at high redshift, would be valuable in order to better characterize the cluster lens population.

\subsubsection{X-ray selection effects}

\begin{figure}[t!]
\begin{center}
  \includegraphics[width=\hsize]{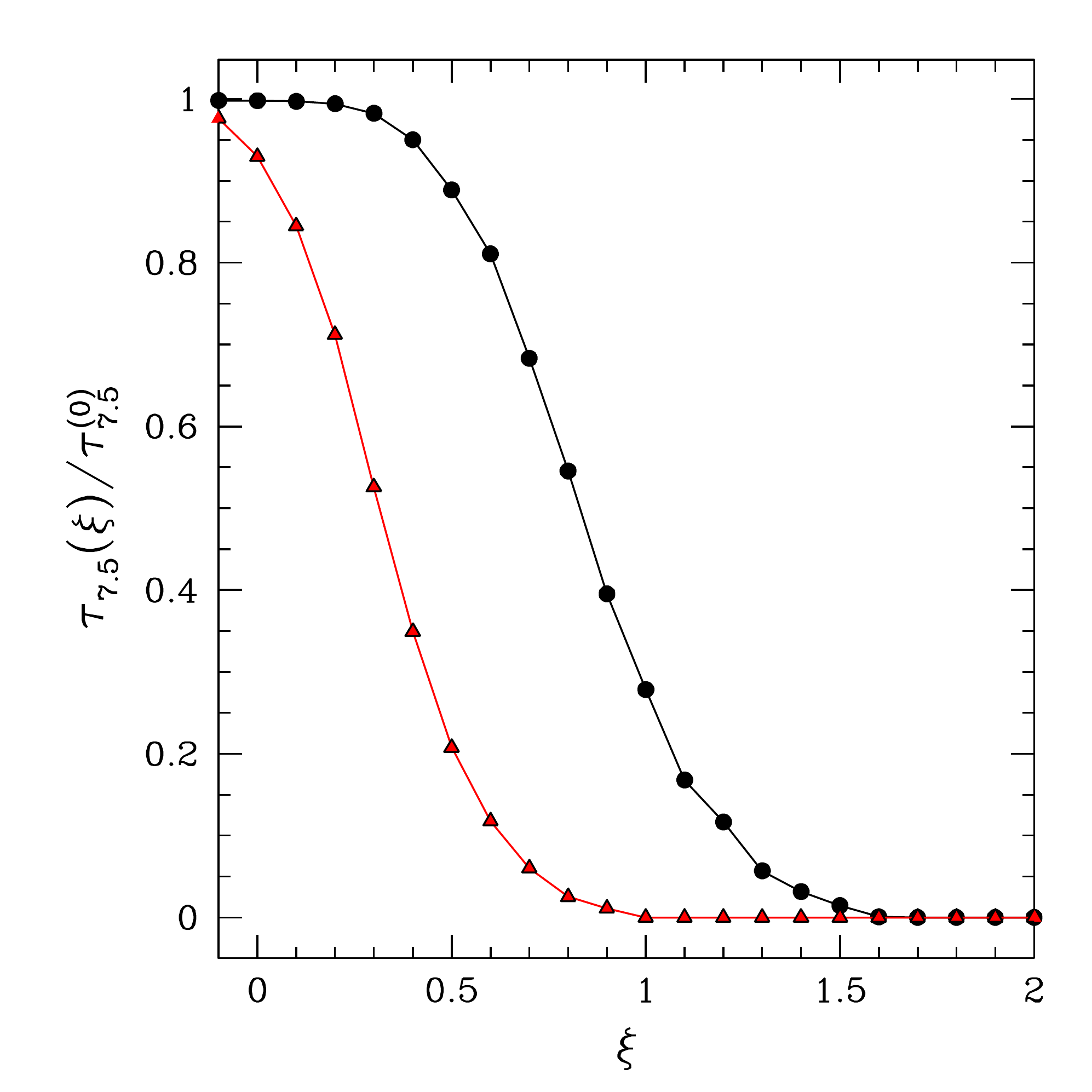}\hfill
\end{center}
	\caption{The fraction of the total optical depth that is contributed by clusters of the {\sc MareNostrum Universe} with logarithm of the bolometric X-ray luminosity (in units of $10^{44}$ erg s$^{-1}$, $\xi = \log[L_\mathrm{X}\,\mathrm{s}/(10^{44}\mathrm{erg})]$) larger than the value reported on the abscissa (filled circles connected by the black line). The filled triangles connected by the red line show the same quantity evaluated for the luminosity in the energy band $[0.5,2]$ keV.}
\label{fig:lumnTau}
\end{figure}

It is finally interesting to investigate how the optical depth changes if clusters are selected with X-ray luminosity. This is particularly worthy since past observational arc statistics studies focused mainly on X-ray selected clusters \citep{GI94.1,LE94.1,LU99.1}. In Figure \ref{fig:lumnTau} we show the fraction of the total optical depth for giant arcs that is contributed by clusters with luminosity above a given threshold. Under the assumption that all sources lie at $z_\mathrm{S} = 2$ this corresponds to the fraction of gravitational arcs that are found inside luminous clusters. Two lines are shown in Figure \ref{fig:lumnTau}, one referring to the bolometric X-ray luminosity (labeled $L_\mathrm{X}$), the other to the luminosity computed in the $[0.5,2]$ keV energy band (labeled $L_\mathrm{B}$). The latter was estimated from the integrated luminosity by using the bolometric correction of \cite{BO99.1}. For that we needed the global cluster temperature, which we estimated from the redshift and the virial mass of the object by using the virial relation 

\begin{equation}
kT = \frac{1.38 \mathrm{keV}}{\gamma}\left( \frac{M}{10^{15}M_\odot h^{-1}} \right)^{2/3} \left[ \Omega_{\mathrm{m},0} \Delta_\mathrm{v}(z_\mathrm{L}) \right]^{1/3}(1+z_\mathrm{L}),
\end{equation}
where $\Delta_\mathrm{v}(z_\mathrm{L})$ is the virial overdensity of collapsed top-hat density perturbations at redshift $z_\mathrm{L}$ and the parameter $\gamma$ defines the ratio of the kinetic energy per unit mass of the dark matter to that of the gas. For simplicity we adopted $\gamma = 1$ henceforth \citep{EV08.1}.

Since no clusters with $L_\mathrm{X} \le 10^{44}$ erg s$^{-1}$ have non-vanishing cross section (see Figure \ref{fig:lsigma}), the totality of the strong lensing optical depth is produced by clusters with bolometric luminosity above that value, as can be seen in Figure \ref{fig:lumnTau}. In other words, if one considers a complete cluster sample containing all the objects with $L_\mathrm{X} > 10^{44}$ erg s$^{-1}$, then this sample would include all the strong lensing clusters in the surveyed area. While several clusters with $L_\mathrm{X} > 10^{45}$ erg s$^{-1}$ exist, their contribution to the total optical depth is small, since they are likely to be very massive and hence extremely rare. As a matter of fact, Figure \ref{fig:lumnTau} shows that such objects contribute only $\sim 20\%$ of the total giant arc abundance. If one considers the band luminosity, clusters with $L_\mathrm{B}>10^{44}$ erg s$^{-1}$ also contribute more than $90\%$ of the total arc abundance. However, the drop of the optical depth with luminosity is somewhat steeper than before, with clusters having $L_\mathrm{B} > 3\times 10^{44}$ erg s$^{-1}$ contributing only $\sim 20\%$ to the total optical depth.

These results are particularly relevant in order to establish the reliability of observational studies directed toward arc statistics. As a classic example (and without the pretense of being rigorous), we considered the work of \citet{LE94.1}, where the authors performed a search for giant arcs inside clusters extracted from the Einstein Medium Sensitivity Survey (EMSS, \citealt{GI94.1}) by having luminosities in the $[0.3,3.5]$ keV energy band larger than $4 \times 10^{44}$ erg s$^{-1}$. Although we did not compute the luminosities in that energy band for all of our simulated clusters , the relative result will lie somewhere in between the red and black lines in Figure \ref{fig:lumnTau}, implying that at least $\sim 20\%$ of the arcs were missed in that sample. We estimated the true luminosity in the $[0.3,3.5]$ keV energy band only for a subsample of the {\sc MareNostrum Universe} lensing clusters, and found that this fraction should be about $30-40\%$. Hence, these observational estimates of arc abundances would have to be corrected for this factor, thus worsening the disagreement with theoretical estimates \citep{FE08.1}. Nevertheless, such a correction is likely to be small compared to other uncertainties concerning theoretical and observational estimates of arc statistics. 

This kind of study is somewhat similar to the analysis performed in \cite{FE07.1}, where the semi-analytic optical depth actually observed in flux limited X-ray cluster samples was detailed. However, the two works cannot be directly compared because there a full account for instrumental effects was presented that is missing here and a different plasma model is used (\citealt{RA77.1}).

\section{Summary and discussion}\label{sct:summary}

In this work we investigated the optical depth emerging from the strong lensing properties of a very large set of numerical galaxy clusters extracted from the {\sc MareNostrum Universe} cosmological simulation. The simulation has the same boxsize of the {\sc Millennium} run \citep{SP05.1}, and includes an equal number of particles for both dark matter and adiabatic gas. The cosmological parameters used in the simulation are in agreement with the WMAP-1 year data release, and in particular the normalisation of the matter power spectrum is $\sigma_8 = 0.9$. Although this is higher than the results of the WMAP-5 year data analysis (\citealt{KO09.1}, see also \citealt{KO10.1}), the two are still consistent at $95\%$ Confidence Level. Additionally, studies of the present abundance of X-ray clusters indicate a somewhat higher normalization than found by the WMAP team \citep{YE07.1,WE10.1,WO10.1}.

Either way, a lower normalization of the matter power spectrum would imply a later formation of cosmic structure, with consequent lower abundance of massive galaxy clusters and a lower average concentration thereof. Both of these effects would act to reduce the optical depth \citep{FE08.1}, however, we expect the \emph{relative} contributions to the arc abundance given by unrelaxed/luminous clusters to be mildly affected. Additionally, a lower dark matter halo concentration might be compensated by the introduction of gas cooling, that would increase the amount of mass flowing to the central region of galaxy clusters.

Before summing up, there is one point of our analysis that we would like to stress. The population of simulated dark matter halos capable of producing giant arcs has a lower mass limit due to the fact that when caustic curves become too small compared to the average source size, the related images can only very rarely be strongly distorted. For this reason isolated massive galaxies, that yet can produce arcs in particularly favorable circumstances, did not enter in the computation of the optical depths. Similarly, the simulation cannot realistically account for the formation of individual galaxies inside clusters due to the lack of baryon cooling, and therefore giant arcs that are mainly produced by those galaxies are also not included in our computations. However we argue for the influence of these issues on the optical depths to be small, since the effect of individual galaxies on the strong lensing cross sections of clusters has already been estimated \citep{ME00.1,WA08.1} to be mild, and the probability of isolated galaxies for producing giant arcs is much too small for it being compensated by their larger abundance. Besides, real strong cluster lensing observations do not focus, by definition, on lensing by isolated galaxies. Nevertheless, the aforementioned limitations of our analysis are to be kept in mind.

The main conclusions that we reached in this paper can be summarized as follows.

\begin{itemize}
\item [$\bullet$] We quantified the correlation between the virial mass of galaxy clusters and the cross section for giant arcs. This relation is approximately linear in logarithm with a slope around  $\sim 1.5$ and accounts for the expected fact that more massive objects have an higher probability of producing strong lensing events. The parameters of the best fit are roughly constant with redshift, although significant oscillations around the mean are displayed.
\item [$\bullet$] 
A similar correlation (with a similar slope) was found between the bolometric X-ray luminosity and the lensing efficiency. This correlation appears to the tighter than the previous one, reflecting the fact that both X-ray emissivity and lensing cross sections are particularly sensitive to the processes occurring at the very center of galaxy clusters.
\item [$\bullet$] Both the correlations described above display a substantial scatter. In particular, the scatter around the best fit mass-cross section relation can be up to one order of magnitude. We found that the distributions around the best fit logarithmic relations are approximately Gaussian, although significant deviations from Gaussianity might be found.
\item [$\bullet$] Using the best fit correlations between mass and lensing efficiency we managed to construct a synthetic optical depth that with only one adjustable parameter manages to reproduce acceptably well the true differential optical depth.
\item [$\bullet$] We computed the contribution to the total optical depth that is given by relaxed and unrelaxed structures, where relaxation was defined as deviation from virial equilibrium. We found that unrelaxed structures mainly contribute to the high-redshift part of the optical depth, in agreement with the standard paradigm for structure formation.
\item [$\bullet$] Similarly, we evaluated the contribution to the total arc abundance given by structures of various X-ray luminosity, thus quantifying the bias present in arc statistic studies based on X-ray selected clusters.
\end{itemize}

Recently \citet{HO10.1} have shown that X-ray selected clusters display a substantially higher giant arc incidence than optically selected clusters with comparable optical luminosity. They conclude that X-ray observables trace much larger mass concentrations than optical ones. Our finding of a relatively tight correlation between the lensing efficiency and the bolometric X-ray luminosity goes in the same direction, and can help interpret the findings of \citet{HO10.1}.

We believe the large set of strong lensing cross sections computed in this work to have a wide range of future applications beyond those exploited here and in Paper I, since it covers an abundant sample of realistically simulated galaxy clusters. We are currently planning to use cluster observables like the X-ray luminosity to construct realistic past light cones, in order to better evaluate possible selection effects and to better compare the lens population to the overall cluster population.

\acknowledgements{Part of this work has been performed under the Project HPC-EUROPA (RII3-CT-2003-506079), with the support of the European Community - Research Infrastructure Action under the FP6 ``Structuring the European Research Area'' Programme. We acknowledge financial contributions from contracts ASI-INAF I/023/05/0, ASI-INAF I/088/06/0 and ASI 'EUCLID-DUNE' I/064/08/0. We warmly thank F. Pace for help in performing the strong lensing analysis and M. Roncarelli for aid in the computation of X-ray luminosities. We additionally would like to acknowledge M. Bartelmann and L. Moscardini for insightful discussions on the manuscript. The {\sc MareNostrum Universe} simulation has been done at BSC-CNS (Spain) and analyzed at NIC J\"ulich (Germany). GY acknowledges support of MICINN  (Spain) through research grants FPA2009-08958 and AYA2009-13875-C03-02. We are grateful to an anonymous referee for comments that helped improving the presentation of our work.}

\bibliography{./master}
\bibliographystyle{aa}

\end{document}